\documentclass[aps,prd,twocolumn,subscriptaddress,amsmath,amssymb
]{revtex4-2}
\usepackage{graphicx}% Include figure files
\usepackage{dcolumn}% Align table columns on decimal point
\usepackage{bm}% bold math
 \usepackage{booktabs}
\usepackage[caption=false]{subfig}
\usepackage{amsmath}
\usepackage{comment}

\usepackage{mathabx}
\usepackage{calligra}
\DeclareMathAlphabet{\mathcalligra}{T1}{calligra}{m}{n}
\DeclareFontShape{T1}{calligra}{m}{n}{<->s*[2.2]callig15}{}

\usepackage{amssymb}
\usepackage{tabularx}
\usepackage{booktabs}
\usepackage{multirow}  % needed for some tables
\usepackage{xcolor}
\usepackage{slashed}

\usepackage{url}
\usepackage{fontawesome5}
\usepackage{scalerel}
\usepackage{tikz}
\usetikzlibrary{svg.path} 
\definecolor{orcidlogocol}{HTML}{A6CE39}
\tikzset{
  orcidlogo/.pic={
    \fill[orcidlogocol] svg{M256,128c0,70.7-57.3,128-128,128C57.3,256,0,198.7,0,128C0,57.3,57.3,0,128,0C198.7,0,256,57.3,256,128z};
    \fill[white] svg{M86.3,186.2H70.9V79.1h15.4v48.4V186.2z}
                 svg{M108.9,79.1h41.6c39.6,0,57,28.3,57,53.6c0,27.5-21.5,53.6-56.8,53.6h-41.8V79.1z M124.3,172.4h24.5c34.9,0,42.9-26.5,42.9-39.7c0-21.5-13.7-39.7-43.7-39.7h-23.7V172.4z}
                 svg{M88.7,56.8c0,5.5-4.5,10.1-10.1,10.1c-5.6,0-10.1-4.6-10.1-10.1c0-5.6,4.5-10.1,10.1-10.1C84.2,46.7,88.7,51.3,88.7,56.8z};
  }
}

\newcommand\orcidicon[1]{\href{https://orcid.org/#1}{\mbox{\scalerel*{
\begin{tikzpicture}[yscale=-1,transform shape]
\pic{orcidlogo};
\end{tikzpicture}
}{|}}}}

\definecolor{mycolor}{RGB}{0,0,204}
\definecolor{pink}{RGB}{255,0,127}
\usepackage[unicode]{hyperref}
\hypersetup{
  colorlinks   = true,    % Colours links instead of ugly boxes
  urlcolor     = blue,    % Colour for external hyperlinks
  linkcolor    =pink,    % Colour of internal links
  citecolor    = pink      % Colour of citations
}

\usepackage{tabularx}
\usepackage{graphicx}% Include figure files
\usepackage{dcolumn}% Align table columns on decimal point
\usepackage{bm}% bold math

\usepackage{float}
\begin{document}

\preprint{APS/123-QED}

\title{Shifted Hybrid Realization of Non-Minimal Higgs Inflation in Light of ACT DR6 and Planck Data} 
\author{Nadir Ijaz \orcidicon{0009-0007-2972-4511}}%
 \email{nadir.ijaz@phd.unipd.it}%
 \affiliation{Dipartimento di Fisica e Astronomia ``G. Galilei", Universit\`a degli Studi di Padova, via Marzolo 8, I-35131 Padova, Italy}
 \affiliation{INFN, Sezione di Padova, via Marzolo 8, I-35131 Padova, Italy}
 \author{Pirzada \orcidicon{0009-0002-2274-9218}}%
 \email{pirzada@itp.ac.cn}%
 \affiliation{CAS Key Laboratory of Theoretical Physics, Institute of Theoretical Physics, Chinese Academy of Sciences, Beijing 100190, China }
 \affiliation{School of Physical Sciences, University of Chinese Academy of Sciences, No. 19A Yuquan Road, Beijing 100049, China  }%
 \author{Mansoor Ur Rehman \orcidicon{0000-0002-1780-1571}}%
 \email{m.rehman@iu.edu.sa}%
 \affiliation{Department of Physics, Faculty of Science, Islamic University of Madinah, Madinah 42351, Saudi Arabia}%

\begin{abstract}
We investigate non-minimal Higgs inflation in a no-scale-inspired supergravity framework and confront its predictions with the latest CMB constraints from ACT DR6 and \emph{Planck}. Working within a shifted hybrid inflation scenario, we construct an effective single-field description in which the GUT Higgs direction serves as the inflaton after stabilization of the orthogonal scalar fields. We show that the inclusion of the leading nonrenormalizable operator in the superpotential induces a controlled deformation of the Starobinsky attractor, 
allowing the scalar spectral index to be shifted into the range favored by recent ACT and related CMB datasets
while maintaining a small tensor-to-scalar ratio, $r \sim 10^{-3}-10^{-2}$. The resulting inflationary dynamics remain theoretically consistent, with controlled supergravity corrections and sub-Planckian inflaton field values. We perform a detailed numerical analysis of the model parameter space, including reheating and nonthermal leptogenesis, and identify regions that simultaneously satisfy current observational constraints and yield a viable post-inflationary cosmological history.
\end{abstract}

\maketitle
\section{\label{sec:level1}Introduction}

The inflationary paradigm has emerged as the foundation of modern cosmology, 
elegantly addressing the horizon and flatness problems while providing a mechanism for the origin of large-scale structure \cite{Guth:1981, Linde:1982, Albrecht:1982,Mukhanov:1981,Hawking:1982,Starobinsky:1982ee,Guth:1982ec}. By positing a period of rapid exponential expansion in the early universe, inflation not only explains the observed homogeneity and isotropy on cosmic scales but also provides a quantum mechanical mechanism for generating primordial density perturbations, which serve as seeds for galaxy formation and the cosmic microwave background (CMB) anisotropies.  Single-field inflation is still the most researched and predictive framework among the many proposed realizations. Its attractiveness stems from its few assumptions and the fact that it yields distinct, verifiable predictions for the scalar spectral index ($n_s$) and the tensor-to-scalar ratio ($r$), which can be directly compared with accurate measurements of the cosmic microwave background (CMB).

Advancements in CMB measurements have continually refined these predictions, imposing increasingly rigorous constraints on inflationary theories. The Planck 2018 results  \cite{Planck:2018vyg,Planck:2018jri} results set benchmark values, with a $n_s = 0.9649 \pm 0.0042$ (68\% CL) and an upper limit on $r < 0.056$ (95\% CL) when paired with BICEP2/Keck Array data~\cite{BICEP:2021xfz}. More recently, the Atacama Cosmology Telescope (ACT) Data Release 6 
(DR6) has provided high-resolution CMB measurements that lead to a mildly 
higher preferred value of the scalar spectral index when combined with 
Planck data, 
$n_s = 0.9709 \pm 0.0038$~\cite{ACTDR6:2025}. This upward shift becomes slightly more pronounced when the analysis is 
supplemented with DESI Year-1 BAO measurements, yielding 
$n_s = 0.9743 \pm 0.0034$
~\cite{ACTDR6:2025,DESI-DR1-BAO}.
These updates are important for differentiating between inflationary models, as they investigate the form of the inflationary potential and the dynamics of the early universe with remarkable accuracy.

However, recent CMB measurements sharpen the preferred range of the scalar tilt and reveal mild tensions across data combinations. The SPT--3G D1 analysis finds that a CMB-only combination of Planck, ACT, and SPT, denoted the ``CMB--SPA'' likelihood, gives \(n_s = 0.9679 \pm 0.0033\)~\cite{SPT-3G:2025bzu}, close to the original Planck constraint~\cite{Planck:2018vyg,Planck:2018jri}. When DESI DR2 BAO data are added to this combination, the preferred value shifts upward to
\(n_s = 0.9726 \pm 0.0028\)~\cite{SPT-3G:2025bzu,DESI:DR2:2025}. This is consistent with ACT-based combinations, which also favor a slightly higher tilt when combined with DESI BAO~\cite{ACTDR6:2025,DESI:DR2:2025}. Importantly, the SPT--3G D1 analysis reports a \(\sim 2.8\sigma\) difference between CMB--SPA and DESI DR2 BAO within \(\Lambda\)CDM~\cite{SPT-3G:2025bzu,DESI:DR2:2025}. Therefore, constraints obtained by combining these data sets should be interpreted with some caution. In what follows, we compare our predictions with both the conservative CMB-only band near \(n_s \simeq 0.968\) and the DESI-informed band near \(n_s \simeq 0.973\)~\cite{Planck:2018vyg,Planck:2018jri,ACTDR6:2025,SPT-3G:2025bzu,DESI:DR2:2025}.

On the theoretical side, plateau or attractor models provide some of the most successful and predictive realizations of inflation. Prominent examples include the $\mathcal{R}^2$ (Starobinsky) model \cite{Starobinsky:1980te} and its superconformal and no-scale supergravity extensions, which generically yield
\begin{equation}
n_s \simeq 1- \frac{2}{N}+{\cal O}\left(\frac{1}{N^2}\right),
\qquad
r \simeq {\cal O} \left(\frac{1}{N^2}\right).
\label{staro_predic}
\end{equation}
For a typical number of e-folds $N\simeq55$, these models predict $n_s\simeq0.964$ and $r\simeq4\times10^{-3}$ \cite{Kallosh:2013hoa,Kallosh:2013}. The same attractor behavior also emerges in non-minimal Higgs inflation and in a broad class of $\alpha$-attractor models, making these predictions remarkably robust across diverse theoretical frameworks. While this universality is one of the major theoretical successes of
attractor inflation, it also motivates the exploration of controlled
departures from the strict attractor limit. Such deformations can arise from higher-dimensional operators, corrections to the Kähler potential, or small departures from exact no-scale structure in supergravity embeddings. Several supersymmetric realizations of Starobinsky-like inflation
\cite{Pallis:2025bzj,Ellis:2025zrf,Ellis:2025ieh,
Pallis:2025gii,Pallis:2025vxo,Ahmed:2026gqq,
Ellis:2026ceb,Pirzada:2026uak}
have recently been proposed to reconcile attractor-based inflationary
predictions with the latest observational data. For representative non-supersymmetric realizations aimed at accommodating
recent observational results, see
\cite{Wolf:2025ecy,Ahmed:2025rrg,Alexandre:2025ixz,
Fu:2025ciy,Kallosh:2025ijd,Ellis:2025bzi,Ahmed:2026msg}. A particularly attractive feature of these modifications is that they can increase the scalar spectral index while preserving the characteristic prediction of a small tensor-to-scalar ratio, typically in the range $r\sim {\rm few}\times10^{-3}-10^{-2}$. 
Interestingly, this is precisely the regime that forthcoming CMB
polarization experiments such as LiteBIRD and CMB-S4 are expected
to probe with unprecedented sensitivity \cite{LiteBIRD:2023,CMB-S4:2019}. These considerations provide a strong motivation for studying well-controlled deformations of attractor inflation within realistic particle-physics frameworks.

Supersymmetric hybrid inflation \cite{Dvali:1994ms,Copeland:1994vg} provides one of the most compelling realizations of inflation within particle physics and grand unified theories \cite{Senoguz:2003zw}. In its standard version with minimal canonical K\"ahler potential, inflation is driven by a gauge-singlet superfield $S$, whose nearly flat potential is generated by supersymmetry and lifted by radiative, supergravity corrections and soft-supersymmetric breaking terms \cite{Senoguz:2004vu,Rehman:2009nq,Rehman:2025fja}. More recently, smooth hybrid inflation has been shown to accommodate the larger values of the scalar spectral index preferred by ACT- and SPT-informed datasets \cite{Okada:2025lpl,Ahmed:2025eip}.
Depending on the structure of the K\"ahler potential, the resulting inflationary predictions can vary significantly \cite{urRehman:2006hu,Rehman:2010wm}. In particular, logarithmic K\"ahler potentials inspired by no-scale supergravity naturally generate non-minimal couplings between the Higgs sector and gravity, thereby realizing non-minimal Higgs inflation within the supersymmetric hybrid inflation framework \cite{Einhorn:2009bh,Ferrara:2010in,Pallis:2011gr,Abid:2021jvn}. In the large-$\xi$ limit, the Einstein-frame potential approaches the
Starobinsky attractor form, yielding the characteristic predictions
displayed in Eq.~(\ref{staro_predic}). While these predictions remain in excellent agreement with the original \emph{Planck} results, the recent preference of ACT- and DESI-combined datasets for a slightly larger scalar spectral index motivates the exploration of well-motivated deviations from the
attractor predictions.

In this paper, we investigate such a departure within the shifted realization of supersymmetric hybrid inflation. Unlike the standard hybrid inflation scenario, where the gauge symmetry remains unbroken during inflation and is subsequently broken at the waterfall transition, shifted hybrid inflation incorporates higher-dimensional superpotential operators that displace the inflationary trajectory away from the origin, allowing the gauge symmetry to be broken already during inflation \cite{Jeannerot:2000sv,Khalil:2010cp,Civiletti:2011qg}. This modification not only avoids the production of topological defects at the end of inflation but also introduces additional flexibility in the inflationary dynamics. By embedding the shifted hybrid inflation framework into non-minimal Higgs
inflation with a logarithmic no-scale K\"ahler potential, we show that the
resulting deformation of the Starobinsky attractor can naturally raise the
scalar spectral index into the range favored by ACT DR6 and related data combinations while maintaining a tensor-to-scalar ratio in the experimentally accessible range $r\sim10^{-3}-10^{-2}$. We perform a comprehensive numerical analysis of the full Einstein-frame dynamics, including reheating and leptogenesis, and identify the regions of parameter space that simultaneously satisfy current observational constraints and provide a realistic cosmological history.

The remainder of this paper is organized as follows. In Section~\ref{Sec2}, we introduce the model and its theoretical framework. In Section~\ref{Sec-III}, we develop its effective single-field realization and derive analytical expressions for the inflationary observables in the slow-roll approximation. In Section~\ref{experiments}, we present the numerical analysis and discuss the regions of parameter space consistent with current observational constraints. Reheating and leptogenesis are studied in Section~\ref{reheating}. Finally, our conclusions are summarized in Section~\ref{conclusion}.

%%%%%%%%%%%%%%%%%%%%%%%%%%%%%%%%%%%%%%%%%%%%%%%%%%%%%%%%%%%%%%%%%%%%%%%%%%%

\section{\label{Sec2}R-Symmetric Higgs Inflation}
In order to investigate non-minimal Higgs inflation, we focus on a specific and phenomenologically realistic example built on the Pati–Salam gauge group $4_c \times 2_L \times 2_R \equiv $ SU(4)$_c$ × SU(2)$_L$ × SU(2)$_R$  \cite{Pati:1973uk,Melfo:2003iw,Pati:1974yy}. In supersymmetric constructions \citep{Khalil:2010cp,Lazarides:2020zof,Lazarides_2020}, shifted 
$\mu$-hybrid inflation is realized through the following superpotential:

\begin{equation}\label{superpotential}
\begin{split}
W = \kappa S (\widebar{H^c} H^c - M^2) + \beta S  \frac{(\widebar{H^c} H^c)^2}{m_P^2} + \lambda S \mathfrak{h}^2 \\
+ a G H^c H^c + b G \widebar{H^c} \widebar{H^c} + \lambda_{ij} F^c_i F_j \mathfrak{h} + (\gamma_{ij}^1 F^c_i F^c_j \\ 
+ \gamma_{ij}^2 F_i F_j) \frac{H^c H^c}{m_P} + (\tilde{\gamma}_{ij}^1 F^c_i F^c_j + \tilde{\gamma}_{ij}^2 F_i F_j) \frac{\widebar{H^c} \widebar{H^c}}{m_P},\end{split}
\end{equation} 
where \(\kappa, \lambda, \beta, a, b, \lambda_{ij}, \gamma_{ij}^1, \gamma_{ij}^2, \tilde{\gamma}_{ij}^1, \tilde{\gamma}_{ij}^2\)  are real and positive dimensionless couplings, $M$ denotes a superheavy mass scale, and, for simplicity, the cutoff scale is taken to be the reduced Planck mass,  $m_P = 2.4\times10^{18}$~GeV. 
The matter and Higgs superfields of the SUSY Pati-Salam model, together with their representations, transformation properties under $4_c \times 2_L \times 2_R$, decompositions under MSSM gauge group $3_c \times 2_L \times 1_Y$, and $U(1)_R$ charge assignments, are summarized in Table~I.

The third term, $\lambda S \mathfrak{h}^2$, yields the effective $\mu$ term. Assuming gravity-mediated SUSY breaking \cite{Chamseddine:1982jx}, the scalar component of $S$ acquires a nonzero vev proportional to the gravitino mass $m_{3/2}$. This induces an effective $\mu$ term, $\mu = -\lambda m_{3/2}/\kappa$, thereby resolving the MSSM $\mu$ problem \cite{Dvali:1997uq}.

The two superpotential terms involving the sextuplet superfield (G) are introduced to generate superheavy masses for $d_H^c$ and $\bar{d}_H^c$,  The $\lambda_{ij}$ terms correspond to the Yukawa couplings and thus generate fermion masses, while the $\gamma_{ij}$ terms give rise to large right-handed neutrino masses required for the see-saw mechanism.

The first three terms in the superpotential ($W$) are relevant for non-minimal Higgs inflation,
\begin{align}
  W \; \supset \; 
  \kappa S (\bar{H}^c H^c - M^2) + \beta S  \frac{(\bar{H}^c H^c)^2}{m_P^2},
\end{align}
for which the global SUSY vacuum is given by
\begin{equation}
\langle S \rangle    = 0, \quad 
\langle \mathfrak{h} \rangle = 0, \quad   v^2 \;\equiv\; \langle H^c \bar H^c\rangle . 
\end{equation}
We note that the nonrenormalizable term proportional to $\beta$ can also appear with other possible structures, such as $(H^c)^4$ and $(\bar{H}^c)^4$.
However, these additional contributions vanish along the $D$-flat right-handed neutrino direction, $( H^c, \, \bar{H}^c ) \supset ( \nu_H^c, \, \nu_H) $, with $|\nu_H^c| = |\nu_H| = h/2$, which is relevant for the subsequent discussion.

The global SUSY vev $v$ of the Higgs fields can be expressed as
\begin{equation}
(v_{\pm}/M)^2 
  =  \frac{-1 \pm \sqrt{ 1 + 4  \xi_{\text{sh}} }}{2
   \xi_{\text{sh}} },
\end{equation}
\begin{figure}[h]
\centering
\includegraphics[width=0.48\textwidth]{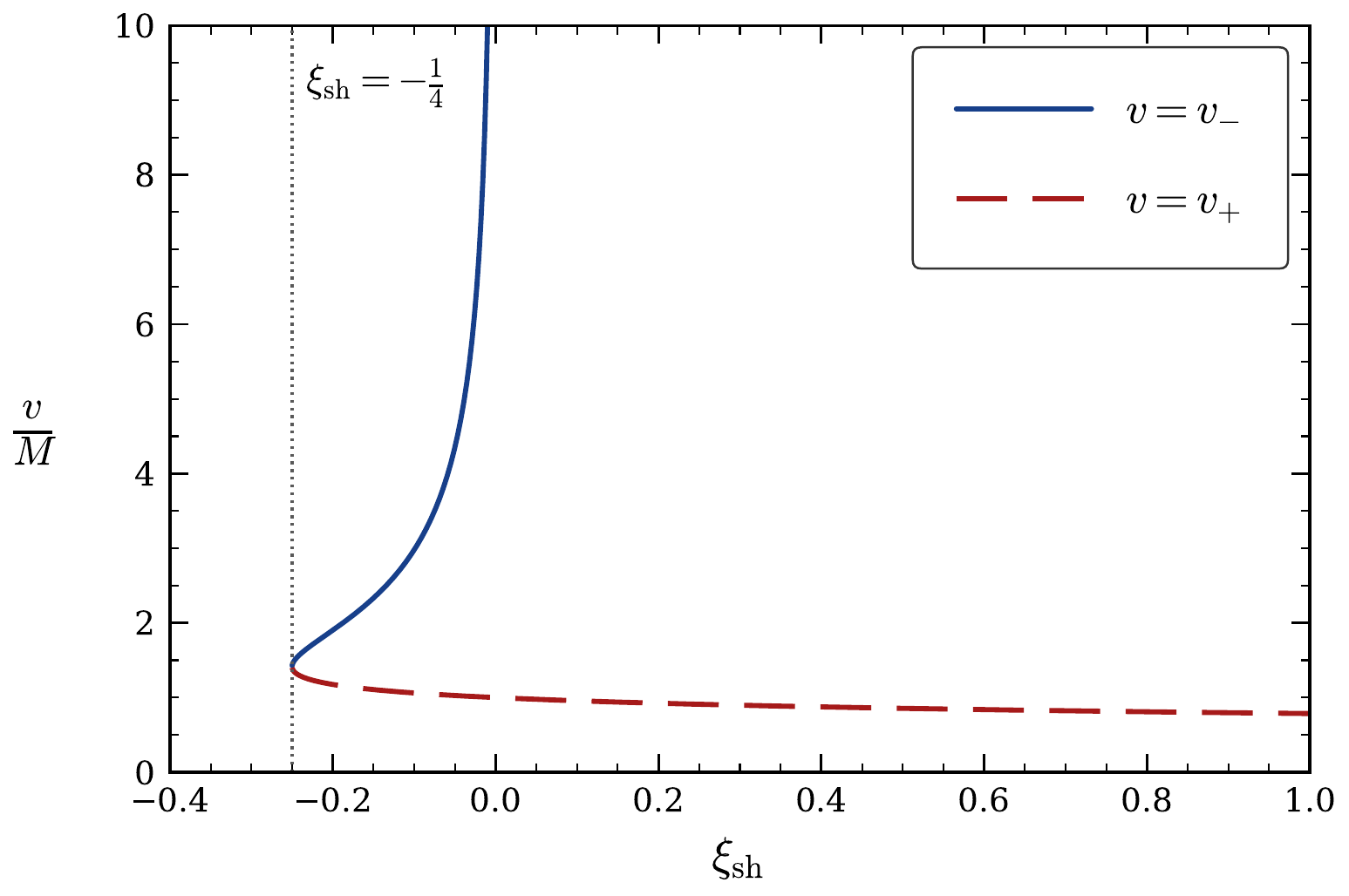} 
\caption{Normalized Higgs vev vs $\xi_{sh}$ parameter. The ratio $v/M$ as a function of the  parameter $\xi_{\rm sh}$. The solid (blue) and dashed (red) curves show the two analytic branches of $v/M$.} 
\label{fig:ac1t}
\end{figure}
where the shifted hybrid inflation parameter $\xi_{\rm sh}$ is defined as
\begin{equation}
\xi_{\rm sh} \equiv \frac{\beta_\kappa M^2}{m_P^2}, \qquad
\beta_\kappa \equiv \frac{\beta}{\kappa}.
\end{equation}
The dependence of the normalized Higgs vev $v/M$ on $\xi_{\rm sh}$ is shown in Fig.~\ref{fig:ac1t}, where the solid and dashed curves correspond to the two analytic branches of the solution. The requirement that $v$ be real imposes the constraint $\xi_{\rm sh} \ge -1/4$, or equivalently,
\begin{equation}
\beta_\kappa \ge -\,\dfrac{m_P^2}{4M^2}.
\end{equation}

This form of the superpotential has previously been employed in shifted $\mu$-hybrid inflation models \cite{Khalil:2010cp,Rehman:2009nq,Rehman:2017gkm,Rehman:2018nsn,Kyae:2005vg,Lazarides:2020zof}, typically with $\beta < 0$. In those scenarios, inflation is driven by the singlet field $S$, while the Higgs fields are stabilized at a local minimum along the $D$-flat direction with $|\nu_H^c| = |\nu_H|$ \cite{Abid:2021jvn}. The parameter range $-1/4 < \xi_{\rm sh} \le -1/8$ is usually considered, corresponding to two distinct solutions for $v$. Beyond its inflationary relevance, the present framework is also connected to other testable aspects of high-scale particle physics. For example, closely related realizations have been shown to predict proton-decay rates that may be accessible to future experiments \cite{Lazarides:2020bgy}.

In contrast, in non-minimal Higgs inflation the roles of the fields are reversed: inflation is driven by the GUT Higgs field $h$, while all other fields, including $S$, are stabilized at the origin. As will be discussed later, we instead focus on the regime $\xi_{\rm sh} > 0$, which allows compatibility with recent CMB observations. For $\beta_\kappa \lesssim 1$ and $M \sim 10^{-2} m_P$, we find $0 < \xi_{\rm sh} \ll 1$, leading to $v \lesssim M$.
In the limit where the nonrenormalizable term is neglected ($\beta = 0$), one obtains $v = M$ \cite{Abid:2021jvn}; however, the resulting predictions for CMB observables ($n_s \lesssim 0.965$ with $N_0  \lesssim 55$) in this limit are in tension with current data.

\begin{table}[ht!]
\centering
\setlength{\tabcolsep}{6pt}
\renewcommand{\arraystretch}{1.15}
\begin{tabular}{l c l c}
\toprule
\textbf{Superfields} & $4_c \times 2_L \times 2_R$ & \multicolumn{1}{c}{$3_c \times 2_L \times 1_Y$} & $q(R)$ \\
\midrule
$F_i$ & $(4,2,1)$ &
\begin{tabular}[t]{@{}l@{}}
$Q_{ia}\,(3,2,\tfrac{1}{6})$ \\
$L_i\,(1,2,-\tfrac{1}{2})$
\end{tabular} & $1$ \\[2pt]
$F^c_i$ & $(\overline{4},1,2)$ &
\begin{tabular}[t]{@{}l@{}}
$u^c_{ia}\,(\overline{3},1,-\tfrac{2}{3})$ \\
$d^c_{ia}\,(\overline{3},1,\tfrac{1}{3})$ \\
$\nu^c_i\,(1,1,0)$ \\
$e^c_i\,(1,1,1)$
\end{tabular} & $1$ \\[2pt]
$H^c$ & $(\overline{4},1,2)$ &
\begin{tabular}[t]{@{}l@{}}
$u^c_{Ha}\,(\overline{3},1,-\tfrac{2}{3})$ \\
$d^c_{Ha}\,(\overline{3},1,\tfrac{1}{3})$ \\
$\nu^c_H\,(1,1,0)$ \\
$e^c_H\,(1,1,1)$
\end{tabular} & $0$ \\[2pt]
$\overline{H^c}$ & $(4,1,2)$ &
\begin{tabular}[t]{@{}l@{}}
$\overline{u^c_{H}}_{a}\,(3,1,\tfrac{2}{3})$ \\
$\overline{d^c_{H}}_{a}\,(3,1,-\tfrac{1}{3})$ \\
$\overline{\nu^c_{H}}\,(1,1,0)$ \\
$\overline{e^c_{H}}\,(1,1,-1)$
\end{tabular} & $0$ \\[2pt]
$S$ & $(1,1,1)$ & $S\,(1,1,0)$ & $2$ \\[2pt]
$G$ & $(6,1,1)$ &
\begin{tabular}[t]{@{}l@{}}
$g_a\,(3,1,-\tfrac{1}{3})$ \\
$g^{a}\,(\overline{3},1,\tfrac{1}{3})$
\end{tabular} & $2$ \\[2pt]
$\mathfrak{h}$ & $(1,2,2)$ &
\begin{tabular}[t]{@{}l@{}}
$\mathfrak{h}_u\,(1,2,\tfrac{1}{2})$ \\
$\mathfrak{h}_d\,(1,2,-\tfrac{1}{2})$
\end{tabular} & $0$ \\
\bottomrule
\end{tabular}
\caption{Superfields together with their decomposition under the SM and their $R$ charge.}
\label{tab:PScontent}
\end{table}

To realize non-minimal Higgs inflation we consider the K\"ahler potential $K$ of the following form:
\begin{equation}
\begin{split}
    K  = &-3 m_P^2\ln\left(1-\frac{1}{3 m_P^2}(|S|^2+|H^c|^2 + |\overline{H^c}|^2) \right.\\& 
    \left. +\frac{\chi}{4m_P^2}(H^c  \overline{H^c} + h.c.) + \frac{\gamma_4}{3m_P^4}|S|^4 \right),
    \end{split}
    \label{p:4}
\end{equation}
where all other fields are assumed to be stabilized at the origin.
Here, $\chi$ and $\gamma_4$ are dimensionless parameters. The quartic term proportional to $\gamma_4$ is introduced to ensure the stabilization of the $S$ field at the origin during Higgs inflation \cite{Senoguz:2004ky, Lee:2010hj}. An alternative approach to stabilizing the $S$ field, without invoking the quartic term in the K\"ahler potential, is to consider a K\"ahler manifold of the form $U(1)_R \times (SU(2)/U(1))_S$ \cite{Pallis:2023gxs}.
It is also worth emphasizing that all terms in the K\"ahler potential preserve the underlying $R$-symmetry. This differs from scenarios in which $R$-symmetry is explicitly broken at the non-renormalizable level in the K\"ahler potential \cite{Ahmad:2025mul}, for example, to facilitate the production of primordial black holes \cite{Ijaz:2023cvc}.

The scalar-gravity portion of the Lagrangian is expressed in the following manner
\begin{equation}
    \mathcal{L}_J = \sqrt{-g_J}\left[\frac{m_P^2}{2}\Omega^2\mathcal{R}_J-g_J^{\mu\nu} \mathcal{G}_{ij} \partial_{\mu}z^i\partial_{\nu}z^{*j}-V_J \right],
\end{equation}
where $\Omega^2=e^{-K/3m_P^2}$ , $\mathcal{R}_J$ is the Ricci scalar in the Jordan frame, $g_J^{\mu\nu}$  is the inverse of the Jordan frame spacetime metric $g^J_{\mu\nu}$, $\mathcal{G}_{ij}$ is the field-space metric in the Jordan frame, $V_J$ represents the scalar potential in the Jordan frame, and  $z_i\in \{H,\,\bar{H},\,S\}$. 
We then proceed with a conformal transformation of the spacetime metric, to work in the Einstein frame
\begin{eqnarray}
    g^E_{\mu\nu} = \Omega^2 \, g^J_{\mu\nu}.
\end{eqnarray}
This transformation leads to the following Lagrangian in the Einstein frame with a canonical coupling to gravity:
\begin{equation}
    \mathcal{L}_E=\sqrt{-g_E}\left[\frac{m_P^2}{2}\mathcal{R}_E-\frac{1}{2}G_{ij}g^{\mu\nu}_E \partial_{\mu}z^i\partial_{\nu}z^{*j} - V_E\right],
\end{equation}
where $G_{ij}$ is the field space metric in Einstein frame and $V_E=\Omega^{-4}V_J$ is the Einstein frame potential. Along the D-flat direction, the supergravity (SUGRA)  scalar potential in the Einstein frame, $V_E$, is defined as,
\begin{equation}
V_E = e^{K/m_P^2}\left((K^{-1})_{ij} D_{z_i}WD_{z^*_j}W^*-3m_P^{-2}\left|W\right|^2\right),
\end{equation}
where,
\begin{equation}
    D_{z_i}W\equiv\frac{\partial W}{\partial z_i}+m_P^{-2}\frac{\partial K}{\partial z_i}W,\quad K_{ij}\equiv\frac{\partial^2K}{\partial z_i\partial z^*_j},
\end{equation}
$D_{z^*_i}W^*=(D_{z_i}W)^*$ and $K$ given in (\ref{p:4}). For the background equations and the relevant power spectra in a multifield framework, see Refs. \cite{Kaiser_2013, Peterson_2011, Gordon_2000, Nakamura_1996, Gong_2011,Geller:2022nkr}.

%%%%%%%%%%%%%%%%%%%%%%%%%%%%%%%%%%%%%%%%%%%%%%%%%%%%%%%%%%%%%%%%%%%%%%%%%%%

\section{\label{Sec-III} Multi-field Analysis}
We can write the complex fields ($S$, $\nu_H$, $\nu^c_H$) in terms of the canonically normalized real scalar fields ($s$, $h$) as 
\begin{equation}
\begin{split}
      S=\frac{1}{\sqrt{2}}s\, e^{i\theta_s},   \quad\nu_H^c \;=\; \frac{h}{\sqrt{2}}\,e^{i\theta_h}\cos\varphi_h,\\\quad
  \overline{\nu^c_{H}} \;=\; \frac{h}{\sqrt{2}}\,e^{i\bar\theta_h}\sin\varphi_h,
  \end{split}
\end{equation}
where $\theta_s$, $\theta_h$ and $\bar\theta_h$ are the respective phases. The D-Flat direction, $ \overline{\nu^c_{H}}= (\nu_H^c)^*$, implies the following conditions,
\begin{equation}
 \varphi_h = \frac{\pi}{4}, \quad \theta_h =  \bar\theta_h.
\end{equation}

A detailed discussion of the stabilization of the fields $s$, $\theta_s$, and $\theta_h$ is in order. For simplicity, we neglect contributions arising from the $\beta$ coupling.
Since inflation is expected to originate at very high energy scales, it is natural to consider Planckian initial conditions. Accordingly, we take the initial values of the scalar fields to satisfy $s \gtrsim m_P \gg M$ and $h \gtrsim m_P$. In the regime where $\chi h^2 \gg m_P^2$, the Einstein-frame scalar potential reduces to \cite{Ijaz:2024zma} 
\begin{eqnarray}
V_E &\simeq&  \frac{4\,\kappa^2 m_P^4}{\chi^2 \cos^2(2\theta_h)} \left[ 1 - \frac{8}{ \cos(2\theta_h)} \frac{m_P^2}{\chi 
 h^2} +  2 \gamma_4 \frac{s^2}{m_P^2} \right.  \nonumber \\
 &+& \frac{16}{3} \frac{s^2}{\chi h^2} \left( \frac{1}{\cos(2\theta_h)} - \cos(2\theta_h)    \right) \nonumber \\
 &+&  \frac{4}{3} \frac{s^2}{\chi h^2} \left. \left( - \cos(2\theta_s) + 3 \tan(2\theta_h) \sin(2\theta_s) \right)  \right]. 
\end{eqnarray}
From this expression, it follows that the potential is minimized at $s = 0$, $\theta_s = 0$, and $\theta_h = 0$, indicating that these field directions are stabilized along the inflationary trajectory.

The exact Einstein-frame potential, expressed as a function of $s$ and $h$, for $\beta_\kappa = 0$ and with all phases stabilized, takes the form
\begin{equation}
V_E(h,s) = \frac{\kappa^2}{16}\,
\frac{\mathcal F}{\Omega^4\,\Delta}
\end{equation}
where,
\begin{equation}
\begin{split}
\Omega^2=&\,1 + \frac{\xi h^2}{m_P^2} - \frac{1}{6} \frac{s^2}{m_P^2}+\frac{\gamma_4}{12} \frac{s^4}{m_P^4},
\\[1mm]
\Delta=&\,1+\xi(1+6\xi) \frac{h^2}{m_P^2} - 2\gamma_4 \frac{s^2}{m_P^2} + \\ & \frac{\gamma_4}{12} \frac{s^4}{m_P^4} - 2\xi(1+6\xi) \gamma_4 \frac{h^2 s^2}{m_P^4},
\\[1mm]
\mathcal F=&\,
(h^2-4M^2)^2\left[1+\xi(1+6\xi) \frac{h^2}{m_P^2}\right]
\\
&-\frac{2}{3}\Bigl[
16M^4
-2\bigl(3 m_P^2 + 4(1+6\xi)M^2\bigr)h^2
\\&+(1+6\xi)h^4
\Bigr]\frac{s^2}{m_P^2}
+\frac{\gamma_4}{12}\Bigl[
400 M^4
\\&-8\bigl(12 m_P^2 + 5(5+24\xi)M^2\bigr)h^2
+(25+144\xi)h^4
\Bigr] \frac{s^4}{m_P^4}
\\& + \frac{\gamma_4}{3} \frac{h^2 s^6}{m_P^4}.
\end{split}
\end{equation}
\begin{figure}[!t]\centering
\includegraphics[width=0.48\textwidth]{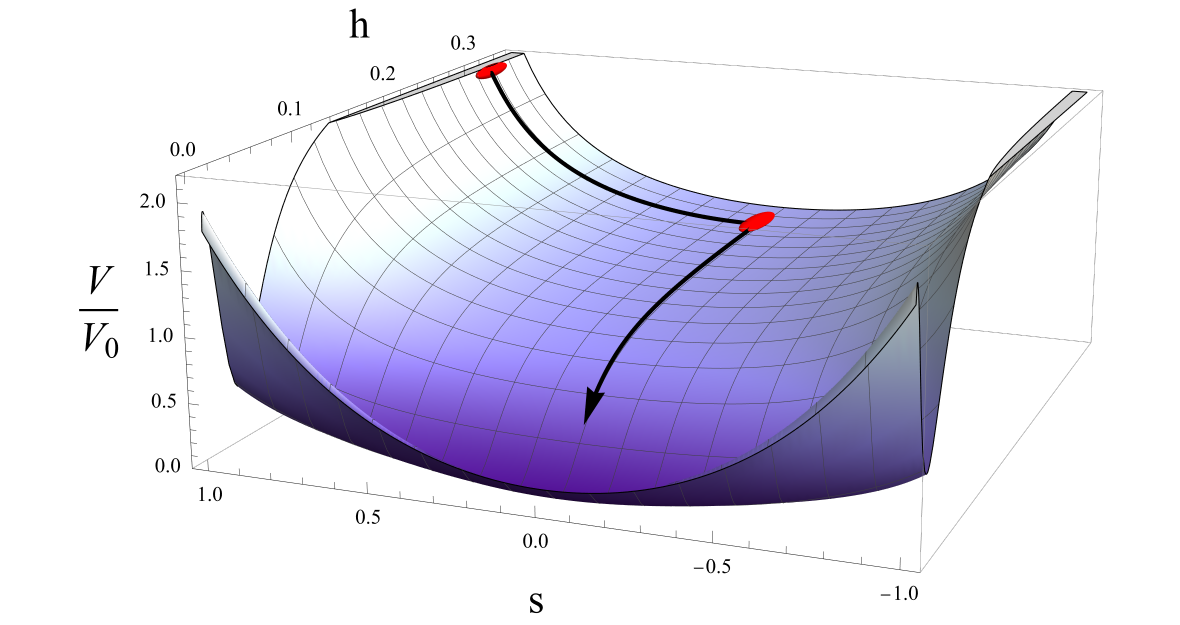}
\caption{\label{fig:3dplot} 
Normalized Einstein-frame potential $V_E/V_0$ with $V_0 = \kappa^2 m_P^4/(16\,\xi^2)$ as a function of $s$ and $h$, for $M = 0.01$, $\xi = 10^3$, $\gamma_4 = 0.3$, and $\beta_\kappa = 0$. The black line with an arrow indicates the inflaton trajectory along the $h$ direction, with blue dots marking key points.}
\end{figure}
The corresponding potential is illustrated in Fig.~\ref{fig:3dplot}, where $V_E/V_0$ is shown as a function of the real scalar fields $s$ and $h$. The potential exhibits a saddle-like structure, with a pronounced stabilization along the $s$-direction. In particular, the presence of the positive $\gamma_4$ term generates an effective mass for $s$, driving it rapidly to its minimum at $s = 0$.
As a result, the inflationary dynamics effectively reduce to a single-field trajectory along the $h$-direction, as indicated by the black line in Fig.~\ref{fig:3dplot}. The stabilization mechanism is robust, with the mass of the $s$ field scaling approximately as $\sim \sqrt{\gamma_4 \xi h^2}$.

Setting $\beta_\kappa = 0$ simplifies the structure of the potential and highlights the dominant role of $\gamma_4$ in stabilizing the transverse direction. When a nonzero $\beta_\kappa$ is introduced, it induces only small corrections to the potential, and the stabilization of the $s$-field remains intact, as verified through numerical analysis.

\section{\label{Sec-IV}Effective Single Field Inflation}
With an appropriate value of $\gamma_4$, $s$ stabilizes at the origin and the scalar potential in the Einstein frame takes the following form,
\begin{equation}
V_E = \kappa^2 M^4 \frac{\left( 1 - \frac{h^2}{(2 M)^2} - \,\beta_{\kappa} \, \frac{h^4}{(2 M)^4} \left( \frac{M}{m_P }\right)^2 \right)^2 }{\left(1 + \frac{\xi h^2}{m_P^2}\right)^2},
    \label{p:2}
\end{equation}
where  $\xi\equiv\frac{\chi}{8}-\frac{1}{6}$.

After performing the conformal rescaling, the canonically normalized inflaton field $\hat{h}(h)$ in the Einstein frame becomes a function of the original field $h$, with
\begin{equation}
J(h)\equiv\left(\frac{d\hat{h}}{dh}\right)=\sqrt{\frac{1}{\Omega^2}+\frac{3}{2}m_P^2\left(\frac{d\ln\Omega^2}{dh}\right)^2
    }.
\end{equation}
The slow-roll parameters can then be expressed in terms of $h$ as
\begin{equation}
\begin{aligned}
\epsilon(h) & = \frac{m_P^2}{2}\left(\frac{V_E^{\prime}}{JV_E}\right)^2, \, 
\eta(h) = m_P^2 \left(\frac{V_E^{\prime\prime}}{J^2V_E}
-\frac{J^{\prime}V_E^{\prime}}{J^3V_E}\right),   \\
\zeta^2(h) & =
m_P^4
\frac{V_E^{\prime}}{J V_E^2}
\left(
\frac{V_E^{\prime\prime\prime}}{J^3}
-\frac{3J^{\prime}V_E^{\prime\prime}}{J^4}
-\frac{J^{\prime\prime}V_E^{\prime}}{J^4}
+\frac{3(J^{\prime})^2V_E^{\prime}}{J^5}
\right).
\end{aligned}
\end{equation}
where primes denote derivatives with respect to $h$. 

To first order in the slow-roll approximation, the scalar spectral index $n_s$, the tensor-to-scalar ratio $r$ and the running of the scalar spectral index are given by
\begin{equation}
\begin{aligned}
n_s  & \simeq 1 - 6 \, \epsilon(h_0) + 2 \, \eta(h_0),\quad r\simeq 16 \, \epsilon(h_0), \\
\alpha_s & \equiv  \frac{d n_s}{d\ln k}
\simeq
16 \epsilon(h_0) \eta(h_0)
-24\epsilon^2(h_0)
-2\zeta^2(h_0),
\end{aligned}
\end{equation}
evaluated at the field value $h_0$, which corresponds to a given number of e-folds.
The number of e-folds before the end of inflation is
\begin{equation}
    N_0=\frac{1}{\sqrt{2}m_P}\int_{h_e}^{h_0}\frac{J(h)}{\sqrt{\epsilon(h)}}dh,
\end{equation}
where $h_e$ denotes the end of inflation, determined by the condition $\epsilon(h_e) = 1$.
The field value $h_0$ is associated with the pivot scale $k_0$, at which the amplitude of the scalar power spectrum is normalized by Planck \cite{Planck:2018vyg} as
\begin{equation}\label{eq20}
   A_s(k_0)=\left.\frac{V_E(h)}{24\pi^2 m_P^4\,\epsilon(h)}\right|_{h(k_0)=h_0}=2.137\times10^{-9},
\end{equation} 
with $k_0=0.05~\text{Mpc}^{-1}$.

Using the mass matrix defined in \cite{Ijaz:2023cvc}, together with the Einstein-frame potential given in Eq.~\eqref{p:2}, the slow-roll parameters in the large-$\xi$ limit and to leading order in $\beta$ can be approximated as
\begin{equation}
\begin{aligned}
\epsilon & \simeq \frac{4}{3\psi^4}+\frac{2\beta_{\kappa}}{3}{\frac{1}{\xi}}, \quad
\eta  \simeq -\frac{4}{3\psi^2}+\frac{\beta_{\kappa}\psi^2}{3\xi}. \\
\zeta^2 & \simeq \frac{16}{9\psi^4}
+\frac{4\beta_{\kappa}}{9\xi}.
\end{aligned}
\end{equation}
Here, $\psi \equiv \sqrt{\xi}h/m_P$ is a dimensionless field variable. At the pivot scale, its value is approximately $\psi_0 \simeq \sqrt{4N_0/3}$, while at the end of inflation it is given by $\psi_e \simeq (4/3)^{1/4} < \psi_0$.
The corresponding expressions for the scalar spectral index $n_s$, the tensor-to-scalar ratio $r$, and the running of the scalar spectral index $\alpha_s$ are
\begin{equation}
\begin{aligned} \label{ns_r}
n_s&\simeq  1-\frac{2}{N_0} +\frac{8\beta_{\kappa}}{9\xi }{N_0},
\quad
r \simeq \frac{12}{N_0^2}+{\frac{32\beta_{\kappa}}{3\xi}}, \\
\alpha_s & \simeq
-\frac{2}{N_0^2}
-\frac{8\beta_{\kappa}}{9\xi}.
\end{aligned}
\end{equation}
These analytical expressions are valid in the limit
$\beta_\kappa \ll \xi / N_0^2 $, i.e., for sufficiently small $\beta_\kappa$, where terms of order $(\beta_\kappa N_0^2/\xi)^2$ and higher are neglected. They clearly demonstrate that both the scalar spectral index $n_s$ and the tensor-to-scalar ratio $r$ increase with increasing $\beta_{\kappa}$ while the running of scalar spectral index $\alpha_s$ decreases with increasing $\beta_{\kappa}$.
%%%%%%%%%%%%%%%%%%%%%%%%%%%%%%%%%%%%%

\subsection*{ACT, SPT, P-ACT-LB, CMB-SPA+DESI}
\label{experiments}

The latest ACT-DR6  release \cite{ACTDR6:2025}, especially when combined with DESI-DR1 \cite{DESI-DR1-BAO}, has sharpened constraints on the scalar tilt, pushing preferred values of the spectral index to the high side of the Planck 2018 \cite{Planck:2018vyg} result $n_s=0.9651\pm0.0044$. Earlier combined analyses already hinted at an upward shift, e.g. $n_s=0.9683\pm0.0040$ \cite{Efstathiou_2021}. The Planck+ACT+LB, DESI-DR1 combination (P-ACT-LB) now yields $n_s=0.9743\pm0.0034$, differing from the Planck value by $\sim 2\sigma$ and, as emphasized in \cite{ACTDR6:2025}, placing classic plateaus such as Starobinsky $R^2$ inflation \cite{Starobinsky:1980te}, non-minimal Higgs inflation \cite{Bezrukov:2008,Okada:2010jf,Masoud:2019cen,Abid:2021jvn,Ijaz:2023cvc} in the large $\xi$ limit, and many $\alpha$-attractor realizations \cite{Kallosh:2013,Kallosh:2013tua,Kallosh:2014rga,Kallosh:2022ggf}  under pressure. A complementary update from SPT-3G (CMB-SPA) \cite{SPT3G:2025} finds $n_s=0.9679\pm0.0033$, close to Planck, but including DESI DR2 shifts the tilt to $n_s=0.9726\pm0.0028$. Importantly, CMB-SPA and DESI DR2 show $\sim2.8\sigma$ tension in $\Lambda$CDM \cite{LindeChaotic2025}, cautioning against firm model exclusions at the $2\sigma$ level based on combinations of datasets that are themselves in significant tension as discussed by the author of \cite{LindeChaotic2025}. Nevertheless, the potential implications for B-mode targets remain notable: as emphasized by CMB-S4 and LiteBIRD forecasts \cite{LiteBIRD:2020khw,CMB-S4:2019}, the current landscape effectively focuses attention on a small set of benchmark classes Starobinsky, Higgs, $T$-model $\alpha$-attractors (continuous $\alpha$), and Poincaré-disk $\alpha$-attractors \cite{Kallosh:2013,Galante:2014ifa,Kallosh_2015,Carrasco:2015pla} with discrete $\alpha$ whose $(n_s,r)$ predictions span the experimentally relevant discovery space.

%%%%%%%%%%%%%%%%%%%%%%%%%%%%%%%%%%%%%

\section{Exact Numerical Analysis}
\begin{figure}[t]
\centering
\includegraphics[width=0.48\textwidth]{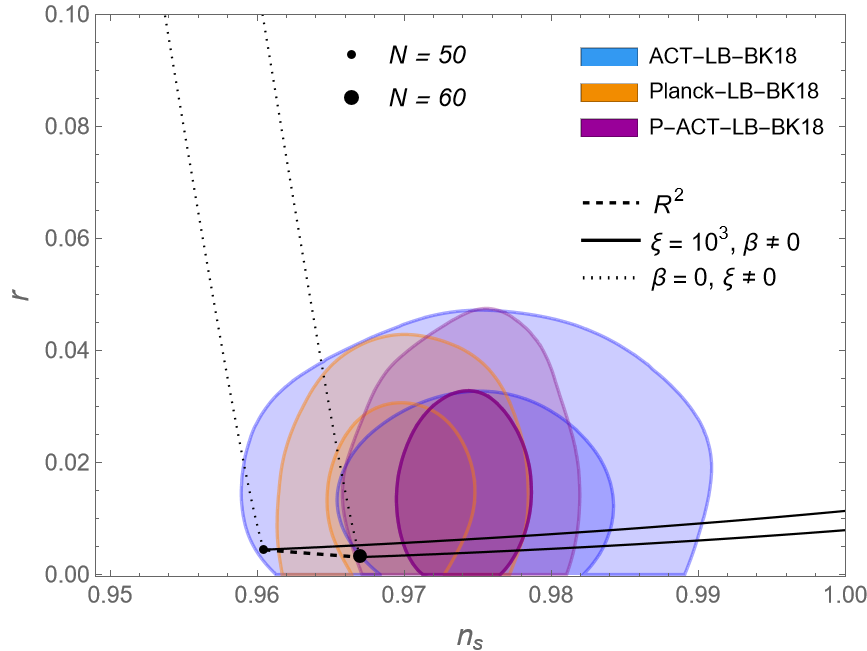} 
\caption{Predictions in the ($n_s,r$) plane together with the latest ACT-LB-BK18, Planck-LB-BK18, and P-ACT-LB constraints. The dashed, dotted, and solid curves correspond to the Starobinsky $\mathcal{R}^2$ model, nonminimal Higgs inflation with $\beta=0$, and nonminimal Higgs inflation with positive $\beta > 0$ and $\xi=10^3$, respectively. In each case, the predictions are displayed for $N_0=50$ and $N_0=60$, assuming a fixed symmetry-breaking scale $M/m_P=0.01$.}
\label{fig:3}
\end{figure}

\begin{figure}[t]\centering
\includegraphics[width=0.48\textwidth]{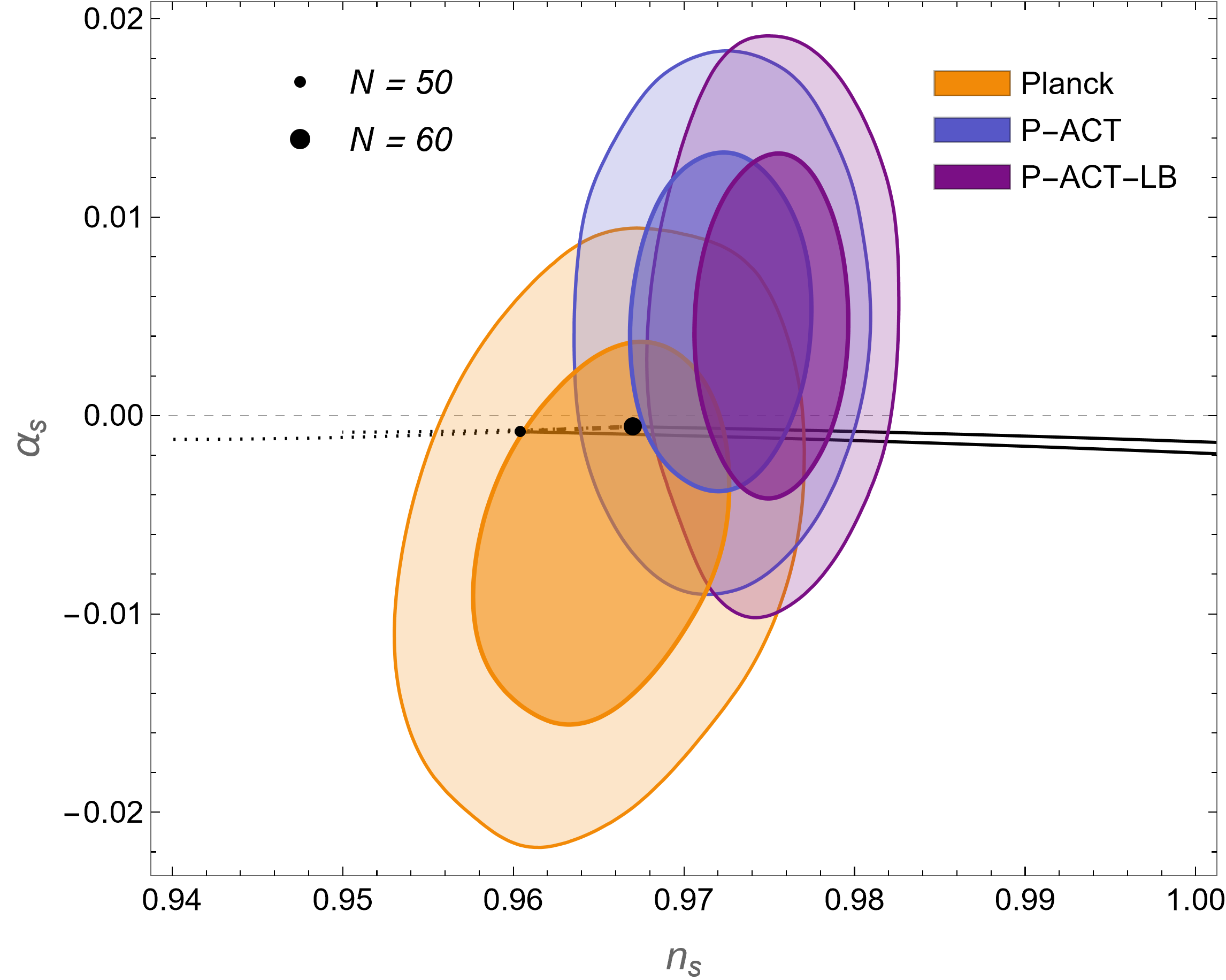}
\caption{  Predicted values of the running of the scalar spectral index, $\alpha_s \equiv dn_s/d\ln k$, in the ($n_s,\alpha_s$) plane. The shaded contours show the marginalized $68\%$ ($1\sigma$) and $95\%$ ($2\sigma$) confidence regions obtained from the Planck (orange), P–ACT (blue), and P–ACT–LB (purple) data combinations. The dashed, dotted, and solid curves correspond to the Starobinsky $\mathcal{R}^2$ model, nonminimal Higgs inflation with $\beta=0$, and nonminimal Higgs inflation with positive $\beta > 0$ and $\xi=10^3$, respectively. In each case, the predictions are displayed for $N_0=50$ and $N_0=60$, assuming a fixed symmetry-breaking scale $M/m_P=0.01$.}
\label{fig:4}
\end{figure}

\begin{figure}[t!]\centering
\includegraphics[width=0.48\textwidth]{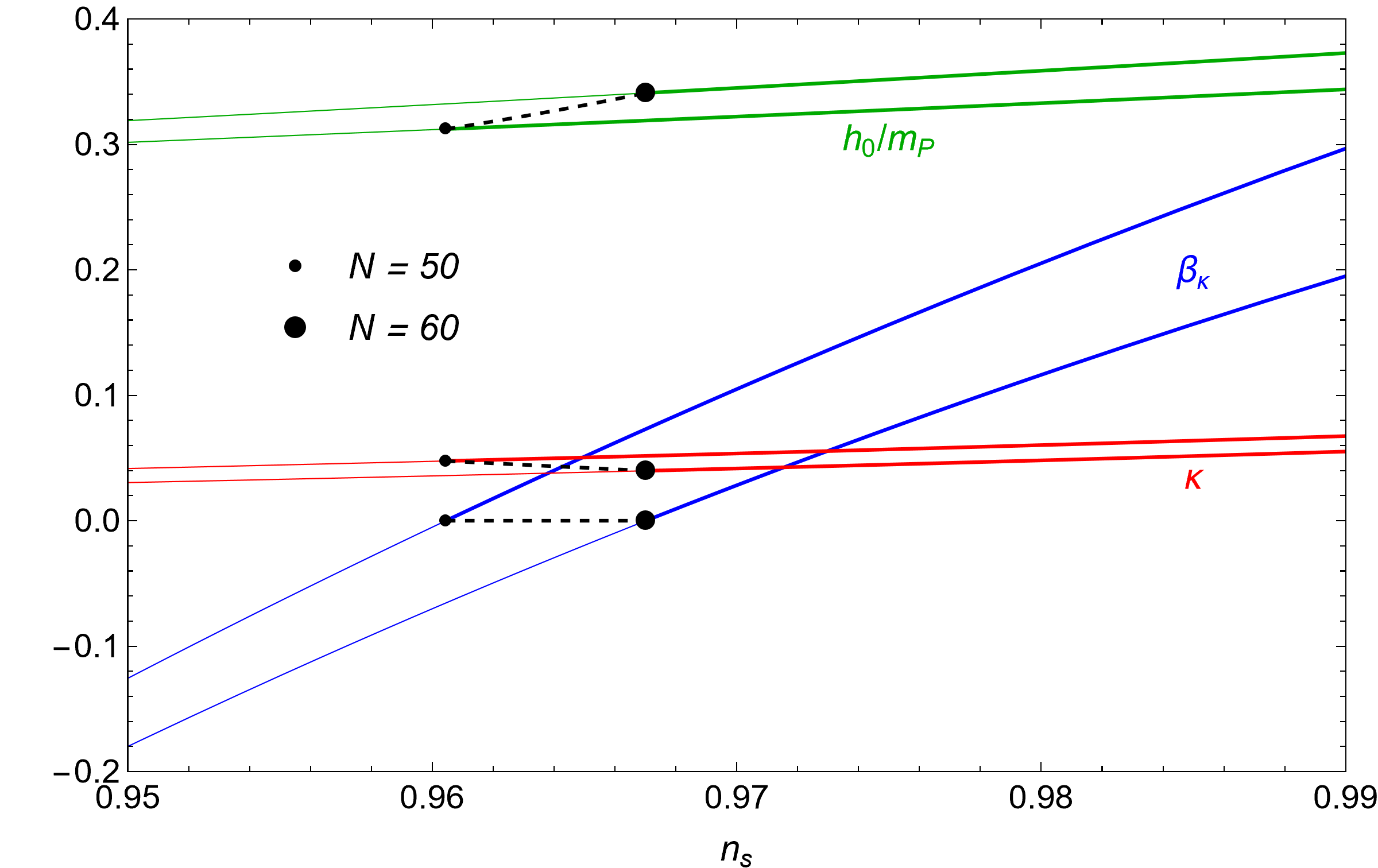}
\caption{Predicted values of the parameters $\beta_{\kappa}$ and $\kappa$, together with the normalized inflaton field value at the pivot scale, $h_0/m_P$, as functions of the scalar spectral index $n_s$ for fixed $\xi = 10^3$ and $M/m_P = 0.01$. The results are shown for $N_0=50$ and $60$ e-folds. The green, blue, and red curves correspond to $h_0/m_P$, $\beta_{\kappa} = \beta / \kappa$, and $\kappa$, respectively, while the markers indicate the predictions in the nonminimal Higgs-inflation limit ($\beta=0$). For each quantity, the thicker and thinner curves correspond to the $\beta >0$ and $\beta <0$ branches, respectively.}
 \label{fig:5}
\end{figure}

We now examine the inflationary predictions of the model and compare them with the latest cosmological constraints. In the limit $\beta=0$, the model reduces to standard nonminimal Higgs inflation. For sufficiently large values of the nonminimal coupling, $\xi \gg 1$, the Einstein-frame potential develops a Starobinsky-like plateau, yielding the universal attractor predictions
\begin{eqnarray}
(n_s, \, r, \, \alpha_s ) &\simeq&  \left( 1-\frac{2}{N_0}, \, \frac{12}{N_0^2}, \, -\frac{2}{N_0^2} \right).
\end{eqnarray}
Numerically, these become
\begin{equation}
\begin{aligned}
(n_s, \, r, \, \alpha_s ) & \simeq   (0.960, 0.005, -8\times 10^{-4})
\text{ for $N_0 = 50$},   \\ 
& \simeq  (0.967, 0.003, -6 \times 10^{-4}) \text{ for $N_0 = 60$}.       
\end{aligned}
\end{equation}
These are the well-known predictions of Starobinsky $\mathcal{R}^2$ inflation and are represented by the dashed curves in Figs.~\ref{fig:3}--\ref{fig:5}. The corresponding inflaton field value at horizon exit,
\begin{equation}
\frac{h_0}{m_P}\simeq
\sqrt{\frac{4N_0}{3\xi}},
\end{equation}
remains sub-Planckian for $\psi_0 \gg 1$ or $\xi\gg N_0$.

In contrast, for $\xi \ll 1$ the conformal flattening becomes ineffective and the model approaches the quartic-potential limit,
\begin{eqnarray}
(n_s, \, r, \, \alpha_s ) &\simeq&  
\left(
1-\frac{3}{N_0},
\frac{16}{N_0},
-\frac{3}{N_0^2}
\right),
\end{eqnarray}
yielding
\begin{equation}
\begin{aligned}
(n_s, r, \alpha_s)
&\simeq
(0.940,0.320,-1.20\times10^{-3})
\text{ for } N_0=50, \\
&\simeq
(0.950,0.267,-8.33\times10^{-4}) \text{ for } N_0=60.
\end{aligned}
\end{equation}
These predictions lie well outside the current CMB bounds, as illustrated by the dotted curves in Figs.~\ref{fig:3} and \ref{fig:4}. The observational contours shown in these figures were obtained from publicly available MCMC chains released through the LAMBDA (Legacy Archive for Microwave Background Data Analysis) archive and analyzed using the \texttt{GetDist} package~\cite{Lewis:2019xzd}. In the quartic-potential limit, the inflaton field value at horizon exit approaches $h_0/m_P\simeq\sqrt{8N_0}$, corresponding to trans-Planckian field excursions. Adopting the conservative requirement $h_0<m_P$ therefore favors values of $\xi$ larger than a few hundred.

The effect of the higher-dimensional operator becomes apparent once $\beta\neq0$. In the large-$\xi$ regime, the additional $h^4$ contribution arising from the shifted-hybrid-inflation sector modifies the shape of the Einstein-frame plateau. The resulting predictions for $n_s$, $r$, and $\alpha_s$ are given analytically in Eq.~(\ref{ns_r}). In particular, the positive correction proportional to $\beta_\kappa$ increases the scalar spectral index while producing only a modest enhancement of the tensor-to-scalar ratio. Consequently, the model naturally accommodates the larger values of $n_s$ favored by the ACT data while maintaining $r$ well below current observational limits. The corresponding predictions for $\xi=10^3$ are shown by the solid curves in Figs.~\ref{fig:3} and \ref{fig:4} for $N_0=50$ and $60$. Values $\beta_\kappa \gtrsim 0.3$ typically lead to $n_s\gtrsim0.99$, which lies outside the preferred ACT-Planck region.

The ACT DR6 analysis~\cite{ACTDR6:2025} allows for a scale-dependent spectral index and finds
\begin{equation}
\alpha_s = 0.0062 \pm 0.0052
\end{equation}
from the P--ACT--LB data combination, consistent with no statistically significant evidence for running. Although Planck alone mildly preferred negative running, the inclusion of ACT shifts the posterior toward slightly positive values while remaining compatible with $\alpha_s \sim 0$. Throughout the phenomenologically viable region of parameter space, our model predicts a small negative running, typically $|\alpha_s|\lesssim10^{-3}$, in agreement with current observational constraints. Moreover, the predicted tensor-to-scalar ratio remains in the range $r\simeq0.003$–$0.01$, making the scenario potentially testable with future CMB B-mode polarization experiments such as LiteBIRD and CMB-S4.

Figure~\ref{fig:5} displays the variation of $\kappa$, $\beta_\kappa$, and the normalized inflaton field value $h_0/m_P$ as functions of $n_s$ for $\xi=10^3$, $M/m_P=0.01$, and $N_0=50$ and $60$. For completeness, both positive and negative values of $\beta$ are shown. Negative values of $\beta$ shift the predictions toward smaller values of $n_s$, thereby worsening agreement with the ACT-preferred region, consistent with the analytical expressions in Eq.~(\ref{ns_r}). Restricting to the $1\sigma$ region of the P--ACT--LB--BK18 data set yields the approximate parameter ranges
\begin{eqnarray}
0.0006 \lesssim \beta \lesssim 0.01,\quad
0.04 \lesssim \kappa \lesssim 0.06, \\
0.01 \lesssim \beta_\kappa \lesssim 0.2,\quad
0.32 \lesssim \frac{h_0}{m_P} \lesssim 0.36.
\end{eqnarray}
These results demonstrate that even a modest higher-dimensional correction can significantly modify the inflationary predictions. Conversely, large values of $\beta_\kappa$ drive the model away from the observationally allowed region, highlighting the importance of nonrenormalizable operators in determining the phenomenology of nonminimal Higgs inflation unless they are suppressed by an underlying ultraviolet completion.

The present framework should be distinguished from the pole-induced Higgs-inflation models of Ref.~\cite{Pallis:2021}, which employ the same shifted-hybrid-inflation superpotential but within a different theoretical setting. In that work, the observationally viable low-$r$ branch is realized by parameterizing the coefficient of the nonrenormalizable superpotential term as $\beta_{\kappa}=1+\delta_{\lambda}$, with $\delta_{\lambda}\sim10^{-5}$, and by exploiting a hyperbolic K\"ahler geometry characterized by a curvature parameter $N$. The role of the higher-dimensional operator in the present model is fundamentally different. Rather than being introduced to cancel the effects of a K\"ahler pole, $\beta_{\kappa}$ modifies the slope and curvature of the Einstein-frame inflationary plateau. As a result, the same class of $R$-symmetric Higgs-sector operators that appears in pole-induced Higgs inflation serves here as the key ingredient for shifting the inflationary predictions toward the higher scalar spectral index favored by the ACT data, while maintaining a tensor-to-scalar ratio within the reach of forthcoming CMB observations.

To sum up, the recent ACT-DR6, Planck, and DESI data favor a higher scalar spectral index, $n_s \simeq 0.9743 \pm 0.0034$ (68\% CL), while maintaining a stringent upper bound $r<0.038$ (95\% CL). These results place mild tension on the Starobinsky $\mathcal{R}^2$ model, whose predictions lie near the lower edge of the preferred $n_s$ region. In contrast, the present model naturally accommodates the higher ACT-preferred values of $n_s$ through the leading nonrenormalizable contribution parameterized by $\beta_\kappa$, while preserving a small tensor-to-scalar ratio within current observational bounds and potentially accessible to future CMB experiments.

%%%%%%%%%%%%%%%%%%%%%%%%%%%%%%%%%%%%%
\subsection*{\texorpdfstring
  Scanning the $\xi$--$\beta_\kappa$ Parameter Space for Compatibility with ACT-DR6 and Complementary Data Sets}
\label{sec:scan}

The analytical results presented in Section~\ref{Sec-III} provide useful insight into the dependence of the inflationary observables on the model parameters. To identify the regions favored by current cosmological data, we performed a comprehensive numerical scan over the nonminimal coupling $\xi$ and the parameter $\beta_\kappa=\beta/\kappa$. Throughout the analysis we fixed the symmetry-breaking scale to $M/m_P=0.01$ and adopted a reheating temperature $T_r=10^8~\mathrm{GeV}$, corresponding to $N_0\simeq53$ e-folds. For each parameter point, the background evolution was solved numerically using the canonically normalized inflaton field, while the coupling $\kappa$ was determined by requiring consistency with the observed amplitude of scalar perturbations at the pivot scale $k_0=0.05~\mathrm{Mpc}^{-1}$. As a theoretical prior, we imposed a sub-Planckian inflaton value at horizon exit, $h_0 \leq m_P$.

\begin{figure}[t]\centering
\includegraphics[width=1.0\linewidth]{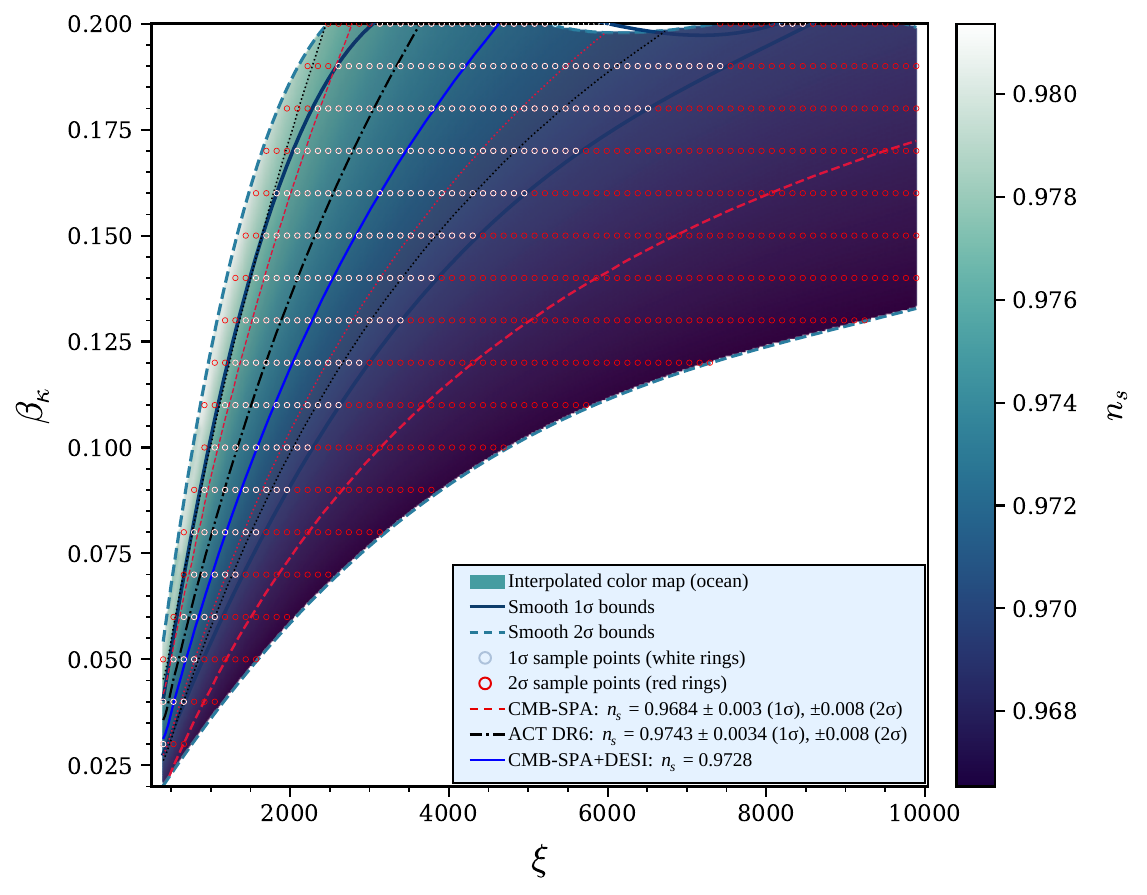} 
\caption{Color–coded map of model predictions in the $(\xi,\beta_\kappa)$ plane, where the color bar indicates the scalar index $n_s$. The interpolated $1\sigma$ (ocean) band is drawn above the $2\sigma$ (rainforest) band for clear separation; unfilled \textbf{white rings} (\textbf{red rings}) mark raw $1\sigma$ ($2\sigma$) samples. External CMB bounds appear as iso–$n_s$ contours with shaded/hatched $\pm1\sigma$ and $\pm2\sigma$ intervals (SPT-(CMB-SPA): dashed crimson; ACT DR6(P-ACT-LB): dash–dot black).  the blue line shows the combine (CMB-SPA+DESI ). All displayed points satisfy the prior $h_0 \le m_P$.}
\label{fig:6}
\end{figure}

\begin{figure}[t]\centering
\includegraphics[width= 1.0\linewidth]{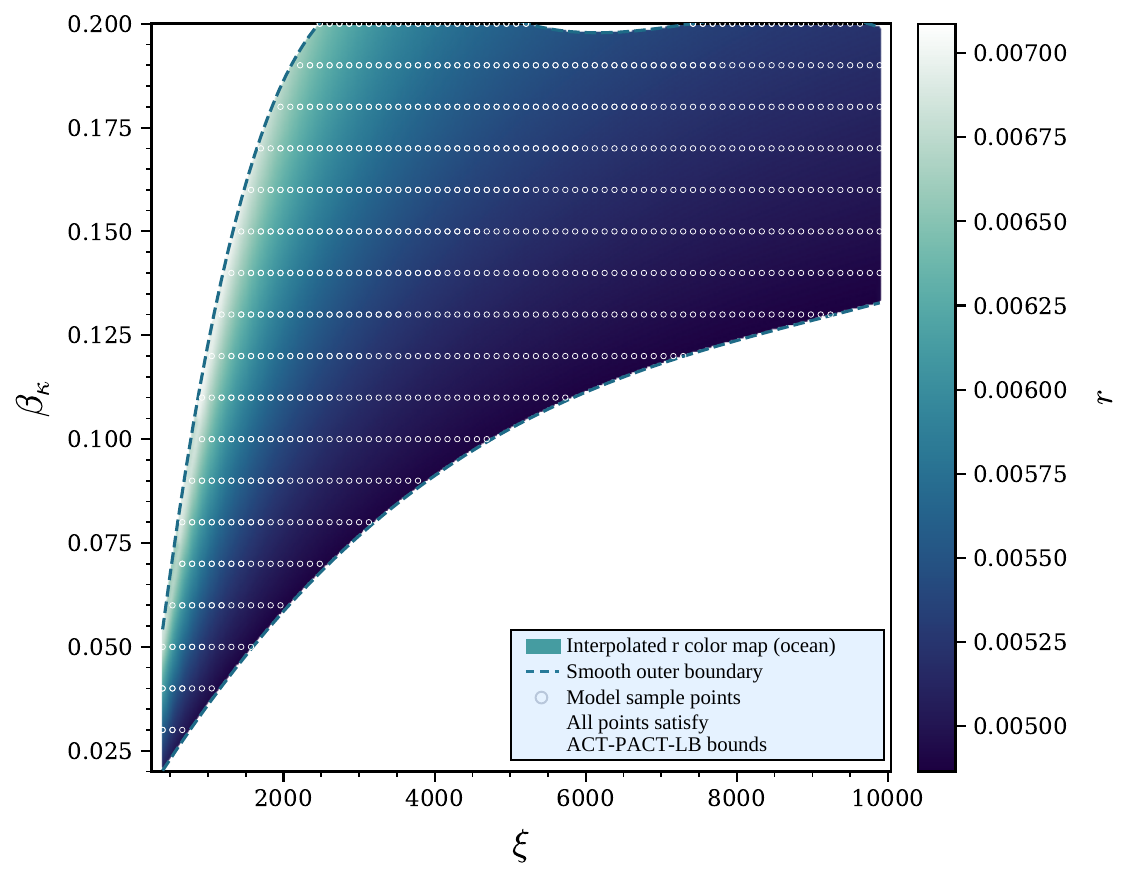}
\caption{Color-coded map of the model predictions in the $(\xi,\beta_\kappa)$ plane, where the color scale represents the tensor-to-scalar ratio $r$. The displayed parameter points satisfy the sub-Planckian field condition $h_0\leq m_P$. See Fig.~\ref{fig:6} for a detailed description of the contours, sample points, and observational constraints.}
\label{fig:7}
\end{figure}

\begin{figure}[t]\centering
\includegraphics[width=1.0\linewidth]{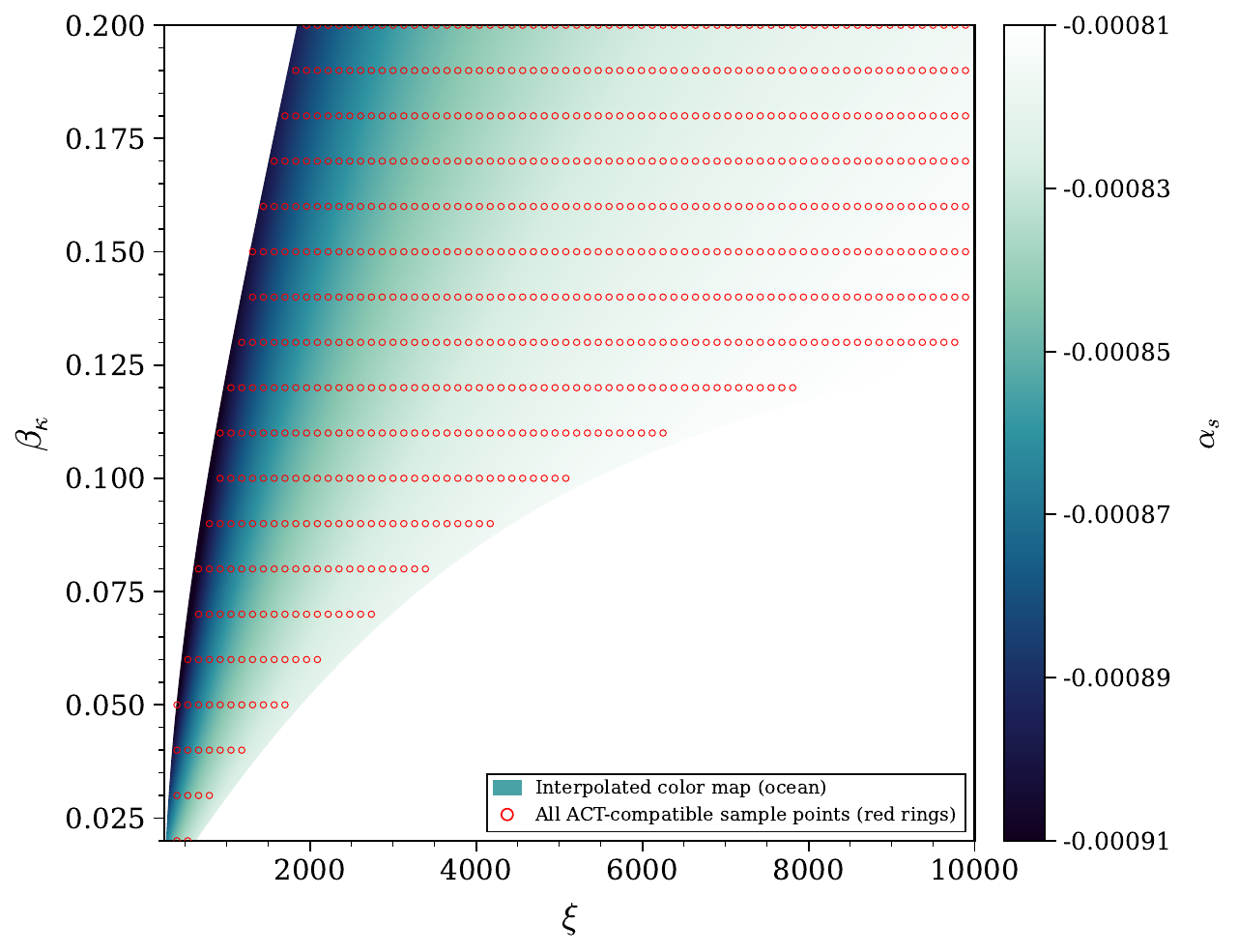} 
\caption{Color-coded map of the model predictions in the $(\xi,\beta_\kappa)$ plane, where the color scale indicates the running of the scalar spectral index $\alpha_s$. The displayed parameter points satisfy the sub-Planckian field condition $h_0\leq m_P$. See Fig.~\ref{fig:6} for a detailed description of the contours, sample points, and observational constraints.}
\label{fig:8}
\end{figure}

The results of the scan are summarized in Figs.~\ref{fig:6}--\ref{fig:8}. Figure~\ref{fig:6} displays the predicted scalar spectral index $n_s$ in the ($\xi,\beta_\kappa$) plane, with each point color-coded according to its $n_s$ value using the \texttt{CMasher} colormap (\texttt{cmr.ocean})~\cite{van_der_Velden_2020}. The white (red) circles denote raw parameter points lying within the $1\sigma$ ($2\sigma$) regions of the scan, while the shaded bands represent smooth interpolations of the corresponding confidence regions. Superimposed contours indicate the preferred values of $n_s$ inferred from the ACT-DR6, Planck, DESI, and CMB-SPA analyses, enabling a direct comparison between the model predictions and current observational constraints. Figures~\ref{fig:7} and \ref{fig:8} show the corresponding predictions for the tensor-to-scalar ratio $r$ and the running of the scalar spectral index $\alpha_s \equiv dn_s/d\ln k$, respectively.

Several important features emerge from the scan. First, increasing $\beta_\kappa$ at fixed $\xi$ systematically shifts the prediction toward larger values of $n_s$, in agreement with the analytical expressions derived in Section~\ref{Sec-III}. This allows the model to interpolate smoothly between the scalar tilt preferred by the CMB-SPA analysis,
\begin{equation}
n_s = 0.9679 \pm 0.0033,
\end{equation}
and the higher value favored by the ACT-DR6+Planck+DESI combination,
\begin{equation}
n_s = 0.9743 \pm 0.0034.
\end{equation}

Second, the parameter combinations consistent with the ACT-preferred region form a broad corridor extending over approximately
\begin{equation}
10^2 \lesssim \xi \lesssim 10^4,
\qquad
0.02 \lesssim \beta_\kappa \lesssim 0.2,
\end{equation}
while maintaining a sub-Planckian inflaton field value. Along this corridor, the coupling constants typically lie in the ranges
\begin{equation}
0.002 \lesssim \kappa \lesssim 0.06,
\qquad
10^{-5} \lesssim \beta \lesssim 10^{-2},
\end{equation}
demonstrating that the required deformation of the inflationary potential originates from a relatively modest higher-dimensional contribution.

The tensor-to-scalar ratio remains small throughout the phenomenologically viable region, as shown in Fig.~\ref{fig:7}. We find
\begin{equation}
0.004 \lesssim r \lesssim 0.007 ,
\end{equation}
well below current observational limits but within the projected sensitivity of future CMB $B$-mode polarization experiments such as LiteBIRD and CMB-S4. Consequently, the model retains the prospect of an observable primordial gravitational-wave signal despite accommodating the larger scalar tilt favored by ACT.

The running of the scalar spectral index, shown in Fig.~\ref{fig:8}, remains negative throughout the allowed parameter space,
\begin{equation}
- 10^{-3}
\lesssim
\alpha_s
\lesssim
-8\times10^{-4},
\end{equation}
with typical values of order $10^{-3}$. These predictions are fully consistent with current observational constraints and constitute a robust prediction of the model.


Overall, the scan reveals a sizable and continuous region of parameter space simultaneously compatible with ACT-DR6, Planck, DESI, and CMB-SPA data. The model successfully accommodates the higher scalar spectral index preferred by ACT through the parameter $\beta_\kappa$, while preserving a small but potentially observable tensor-to-scalar ratio, a negative running of order $10^{-3}$, and sub-Planckian inflaton field values. This demonstrates that the scenario remains both theoretically well-controlled and observationally testable in forthcoming CMB measurements.

%%%%%%%%%%%%%%%%%%%%%%%%%%%%%%%%%%%%%%%%%%%%%%%%%%%%%%%%%%%%%%%%%%%%%%%%%%%
\section{Reheating and leptogenesis}
\label{reheating}
Following the end of inflation, the inflaton field oscillates about the minimum of its potential and transfers its energy to the thermal bath through its decay products, thereby reheating the Universe. In the present framework, an important decay channel is the production of the lightest right-handed neutrino (RHN), which subsequently enables non-thermal leptogenesis. The corresponding partial decay width is given by \cite{Pallis_2011}
\begin{equation}
    \Gamma_{N}=\frac{1}{64\pi}\left[ \frac{M_{N}}{M}\frac{\Omega_{0}^{3/2}}{J_0}\left(1-\frac{12\xi M^2}{M_P^2}\right)\right]^2 m_\text{inf}\sqrt{1-\frac{4M_N^2}{m_\text{inf}^2}},
\end{equation}
where
\begin{equation}
m_{\rm inf}=\frac{\sqrt{2}\,\kappa M}{\Omega_0 J_0}
\end{equation}
denotes the inflaton mass. The quantities
\begin{equation}
J_0 = \sqrt{\frac{1}{\Omega_0}+\frac{6\xi^2M^2}{M_P^2\Omega_0^2}}, \quad
\Omega_0 = 1 + \frac{4\xi M^2}{M_P^2},
\end{equation}
are evaluated at the vacuum, and $M_N$ represents the mass of the lightest RHN. For the parameter region of interest, we typically find
\(
M_N \sim 10^{10}-10^{12},\mathrm{GeV}
\)
and
\(
m_{\rm inf}\sim 10^{12}-10^{13},\mathrm{GeV}.
\)

In addition to RHN production, the inflaton can decay through Yukawa interactions. In supergravity realizations, the dominant contribution arises from the top-Yukawa operator $y_{33}^{u,\nu}Q_3L_3H_u$ \cite{Endo_2006}. For no-scale-like supergravity models, the corresponding decay width is given by
\cite{Pallis:2011gr},
\begin{equation}
    \Gamma_y=\frac{3}{128\pi^3}\left(\frac{6\xi y \Omega^{3/2}_0}{J_0}\right)^2\left(\frac{M}{M_P}\right)^2\left(\frac{m_\text{inf}}{M_P}\right)^2 m_\text{inf},
\end{equation}
where $y\equiv y_{33}$ denotes the top Yukawa coupling.

The reheating temperature is determined by the total inflaton decay width,
\begin{equation}
    T_r=\left(\frac{72}{5\pi^2g_{*}}\right)^{1/4}\sqrt{\Gamma M_P},\quad \Gamma= \Gamma_N+ \Gamma_y,
\end{equation}
with $g_*\simeq228.75$ corresponding to the MSSM particle content. Assuming a standard post-inflationary thermal history, the number of e-folds between horizon exit and the end of inflation can be approximated by \cite{Liddle_2003}
\begin{equation}
N_0 = 53 + \frac{1}{3}\ln\!\left(\frac{T_r}{10^9\,\text{GeV}}\right)
+ \frac{2}{3}\ln\!\left(\frac{\sqrt{\kappa}\,M}{10^{15}\,\text{GeV}}\right).
\end{equation}

For our benchmark choice
\begin{equation}
M=0.01 m_P,\qquad
\kappa=0.0054,\qquad
\xi=10^3,
\end{equation}
we obtain
\begin{equation}
\Omega_0\simeq1.4,\qquad
J_0\simeq17.5,\qquad
m_{\rm inf}\simeq7\times10^{12}\ {\rm GeV}.
\end{equation}
Taking a representative value \(y\equiv y_{33}=0.5\), the resulting RHN decay widths, reheating temperatures, and corresponding values of $N_0$ are summarized below:
\[
\begin{array}{|c|c|c|c|}
\hline
M_N~[\mathrm{GeV}] & \Gamma_N~[\mathrm{GeV}] & T_r~[\mathrm{GeV}] & N_0 \\
\hline
1.0\times 10^{10} & 2.31 \times 10^{-6} & 2.91 \times 10^{8} & 52.960 \\
1.0\times 10^{11} & 2.31\times 10^{-4} & 2.92\times 10^{8} & 52.966 \\
1.0\times 10^{12} & 2.2 \times10^{-2} & 2.98\times 10^{8} & 52.975 \\
\hline
\end{array}
\]
These results indicate a relatively weak dependence of the reheating temperature on the RHN mass over the range $10^{10}\lesssim M_N\lesssim10^{12}$~GeV. We find
$T_r \simeq 2.91-2.98\times10^8~{\rm GeV}$,
while the corresponding number of e-folds remains close to
\(
N_0\simeq52.9
\).
Although the top-Yukawa channel provides the dominant contribution to the total decay width for the benchmark point, RHN production remains sufficiently efficient to generate the observed baryon asymmetry through non-thermal leptogenesis.

An important consistency requirement in supergravity inflationary models is the avoidance of gravitino overproduction \cite{Ellis:1984eq1}. The resulting bound constrains the reheating temperature and depends on the gravitino mass and its stability. For unstable gravitinos with masses above $\mathcal{O}(10)$ TeV, the constraint becomes relatively insensitive to $m_{3/2}$, whereas for stable gravitinos one typically requires $T_r\lesssim10^9$ GeV \cite{Kawasaki_2005,Kawasaki_2008,Bezrukov_2018}. The reheating temperatures obtained above,
$T_r \simeq (2.91-2.98)\times10^8~{\rm GeV}$,
comfortably satisfy these bounds.

The generated RHNs decay out of equilibrium and produce a lepton asymmetry, a fraction of which is subsequently converted into the observed baryon asymmetry through electroweak sphaleron processes \cite{Shaposhnikov:1991pd,Harvey:1990qw,FukugitaYanagida:1986}. We, therefore, investigate the viability of non-thermal leptogenesis \cite{Senoguz:2003zw,Senoguz:2004ky,Senoguz:2004vu}.
The resulting lepton asymmetry is approximately
\begin{equation}
\frac{n_L}{s}
\simeq
\frac{3T_r}{2m_{\rm inf}}
\left(\frac{\Gamma_N}{\Gamma}\right)
\epsilon_1 ,
\end{equation}
where $\epsilon_1$ denotes the CP asymmetry generated in the decay of the lightest RHN. Assuming a hierarchical neutrino spectrum, one obtains \cite{hamaguchi2002cosmological}
\begin{equation}
\epsilon_1
\simeq
-\frac{3m_{N_3}M_N}
{8\pi v_u^2}
\delta_{\rm eff},
\end{equation}
where $v_u$ is the vacuum expectation value of the up-type Higgs field, $m_{N_3}$ is the heaviest light-neutrino mass, and $\delta_{\rm eff}$ parametrizes the effective CP-violating phase \cite{DavidsonIbarra:2002}. Substituting this expression yields
\begin{equation}
\begin{aligned}
\frac{n_L}{s}
&\simeq
b
\left(\frac{\Gamma_N}{\Gamma}\right)
\left(\frac{T_r}{m_{\rm inf}}\right)
\left(\frac{M_N}{10^6{\rm GeV}}\right)
\left(\frac{m_{N_3}}{0.05{\rm eV}}\right)
\delta_{\rm eff}.
\end{aligned}
\label{eq:lepto_compact}
\end{equation}
where $b = 3.9\times10^{-10}$.

Throughout our analysis we adopt $m_{N_3}=0.05$~eV and assume a hierarchical RHN spectrum,
\begin{equation}
M_N \ll M_{N_2} \ll M_{N_3}.    
\end{equation}
The lightest RHN mass is scanned subject to the requirements of non-thermal production,
$M_N>T_r$,
and kinematic accessibility,
$M_N<m_{\rm inf}/2$.
Demanding successful leptogenesis while imposing the theoretical bound $|\delta_{\rm eff}|\le1$, we find the minimal viable solution
\begin{equation}
M_N^\star\simeq9.5\times10^{11}\ {\rm GeV},
\end{equation}
for which
\begin{equation}
\left.\frac{n_L}{s}\right|_{\delta_{\rm eff}=1}
\simeq6.4\times10^{-10},
\end{equation}
in agreement with the observed baryon asymmetry~\cite{ParticleDataGroup:2024cfk}. The corresponding benchmark values are summarized in Table~\ref{tab:lepto_benchmark}. For this benchmark, reproducing the observed asymmetry requires $\delta_{\rm eff}\simeq0.96$,
demonstrating that successful non-thermal leptogenesis can be realized within the allowed parameter space of the model.
\begin{table}[t]
\caption{Non-thermal leptogenesis: Minimal benchmark point satisfying $M_N>T_r$ and $M_N<m_{\rm inf}/2$ with $|\delta_{\rm eff}|\le 1$.}
\begin{ruledtabular}
\begin{tabular}{l c}
Quantity & Value \\
\hline
$M_N^\star$ & $9.5\times10^{11}\ \mathrm{GeV}$ \\
$T_r^\star$ & $2.97\times10^{8}\ \mathrm{GeV}$ \\
$M_N^\star/T_r^\star$ & $3.18\times10^{3}$ \\
$m_{\mathrm{inf}}$ & $7.0\times10^{12}\ \mathrm{GeV}$ \\
$\Gamma_N^\star$ & $2.0\times10^{-2}\ \mathrm{GeV}$ \\
$\Gamma_y^\star$ & $4.42\times10^{-1}\ \mathrm{GeV}$ \\
$\left.n_L/s\right|_{\delta_{\mathrm{eff}}=1}$ & $6.39\times10^{-10}$ \\
$N_0^\star$ & $52.98$ \\
$\delta_{\mathrm{eff}}$ for $n_L/s=6.37\times10^{-10}$ & $0.96$ \\
\end{tabular}
\end{ruledtabular}
\label{tab:lepto_benchmark}
\end{table}

\section{CONCLUSION}
\label{conclusion}
In this work, we have investigated a non-minimally coupled, $R$-symmetric Higgs inflation scenario embedded within the supersymmetric Pati--Salam framework. Employing the full Einstein-frame dynamics and fixing the inflationary scale through the observed amplitude of scalar perturbations, we performed a numerical analysis of the inflationary evolution without relying on slow-roll approximations. The post-inflationary evolution was treated consistently through reheating and leptogenesis, allowing a unified study of inflationary predictions and their cosmological implications.

A comprehensive scan of the $(\xi,\beta_\kappa)$ parameter space reveals an extended region that remains compatible with current cosmological observations. In particular, we find that for
$\xi \sim 3\times10^2-10^5,
\,
\beta_\kappa \sim 0.01-0.2$,
the model naturally predicts a scalar spectral index in the range preferred by the latest ACT-DR6 analyses,
$n_s \simeq 0.971-0.974$,
while yielding a tensor-to-scalar ratio
$r \simeq (4.6-9)\times10^{-3}$
throughout the central region of parameter space. Larger values,
$r \lesssim 2\times10^{-2}$,
are obtained near the boundaries of the observationally allowed region. We further find that the mild differences between the preferred spectral tilts inferred from ACT and SPT data can be accommodated through the intrinsic $\xi$--$\beta_\kappa$ parameter degeneracy, without requiring any modification of the underlying framework.

The reheating dynamics were analyzed by including inflaton decays into right-handed neutrinos and Yukawa-coupled matter fields. For representative benchmark points, the resulting reheating temperature is of order
$T_r \sim 10^8~{\rm GeV}$,
which is consistent with gravitino constraints and leads to a number of e-folds $N_0 \simeq 53$. Incorporating the neutrino sector naturally associated with the Pati--Salam symmetry, we showed that successful non-thermal leptogenesis can be realized while satisfying both the kinematic requirements for inflaton decay and the observed baryon asymmetry of the Universe. The inflationary and baryogenesis sectors therefore remain mutually consistent throughout the phenomenologically relevant parameter space.

Overall, the model provides a coherent and predictive realization of Higgs inflation within a supersymmetric grand unified framework, linking the inflationary epoch, reheating, neutrino physics, and the generation of the cosmic matter-antimatter asymmetry. The predicted tensor-to-scalar ratio lies within the projected sensitivity of upcoming CMB polarization experiments such as LiteBIRD, CMB-S4, and the Simons Observatory. Consequently, the parameter regions identified in this work will be subject to stringent experimental scrutiny in the near future. Improved measurements of primordial gravitational waves, together with progress in neutrino physics and baryogenesis studies, will provide important tests of the scenario and may offer valuable insight into the connection between inflation and grand unification.

\section{Acknowledgments}
NI  acknowledge support from Istituto Nazionale di Fisica Nucleare (INFN) through the Theoretical Astroparticle Physics (TAsP) project. M.R. extends his appreciation to the Deanship of Scientific Research, Islamic University of Madinah, Saudi Arabia, for funding this research work.

\bibliography{NMHI}% Produces the bibliography via BibTeX.
\end{document}